\documentclass[12pt]{article}
\usepackage{epsfig}
\usepackage{amsmath}
\usepackage{hhline}
\usepackage{amssymb}
\usepackage{times}
\usepackage{cite}

\newlength{\dinwidth}
\newlength{\dinmargin}
\def\Journal#1#2#3#4{{#1} {\bf #2}, #3 (#4)}

\def\NPB{{\em Nucl. Phys.} B}
\def\PLB{{\em Phys. Lett.}  B}
\def\PRL{\em Phys. Rev. Lett.}
\def\PRD{{\em Phys. Rev.} D}
\def\ZPC{{\em Z. Phys.} C}
\def\EPJ{{\em Eur. J. Phys.} C}
\def\PR{\em Phys. Rept.}

\begin{document}
\begin{titlepage}

\vspace*{2cm}

\begin{center}
\begin{LARGE}

{\bf Hadronic Structure Functions\footnote{Invited opening plenary talk at the 
8th International Workshop on Deep Inelastic Scattering and QCD, \\
Liverpool, UK (2000)}}

\end{LARGE}

\vspace{1.7cm}
{\large Martin Erdmann}

\vspace{1.2cm}

Institut f\"ur Experimentelle Kernphysik,
Universit\"at Karlsruhe, \\ Engesserstr. 7, 
D-76128 Karlsruhe,
Martin.Erdmann@desy.de

\end{center}

\vspace{2cm}

\begin{abstract}
Experimental results on hadronic structures are discussed 
in view of our physics understanding.
Achievements and challenges are noted. 
\end{abstract}

\end{titlepage}

\section{Motivation}

\begin{figure}[hhb]
\setlength{\unitlength}{1cm}
\begin{picture}(16.0,4.0)
\put(0.0,2.8){\Large LEP/TESLA}
\put(0.1,-0.1){${e}$}
\put(0.1,1.2){${e}$}
\put( 2.1, 0.2){${e}$}
\put( 2.1, 0.9){${e}$}
\put( 0.4,0)
{\epsfig{file=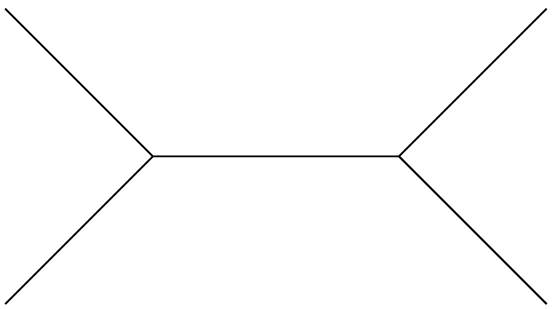,width=1.2cm,angle=90.}} 
\put(2.4,0.3)
{\epsfig{file=feynman.eps,width=1.2cm}} 
\put(5.4,2.8){\Large TEV/LHC}
\put( 5.1,-0.1){${q},{g}$}
\put( 5.1, 1.2){${q},{g}$}
\put( 7.4, 0.2){${q},{g}$}
\put( 7.4, 0.9){${q},{g}$}
\put(5.8,0)
{\epsfig{file=feynman.eps,width=1.2cm,angle=90.}} 
\put(8.1,0.3)
{\epsfig{file=feynman.eps,width=1.2cm}} 
\put(10.5,2.8){\Large HERA}
\put(10.5,2.3){\Large LEP/TESLA}
\put(10.6,-0.1){${q}$}
\put(10.6, 1.2){${e}$}
\put(12.8, 0.2){${q}$}
\put(12.8, 0.9){${e}$}
\put(13.55,0.9){\large ?}
\put(10.9,0)
{\epsfig{file=feynman.eps,width=1.2cm,angle=90.}} 
\put(13.1,0.3)
{\epsfig{file=feynman.eps,width=1.2cm}} 
\end{picture}
\caption{\label {fig:motivation}
Basic diagrams at electron--positron, hadron--hadron, and
lepton--hadron colliders.
Only the last diagram is forbidden within the Standard Model.
}
\end{figure}

\noindent
Today's motivation of measuring lepton--hadron scattering processes
is at least four-fold.
Fig.\ref{fig:motivation} shows basic diagrams at
electron--positron, hadron--hadron, and lepton--hadron colliders:
\begin{itemize}
\item
Only in lepton--hadron collisions is the fusion diagram forbidden
within the Standard Model, which strongly motivates 
\mbox{\bf searches for new}  {\bf physics}, e.g. leptoquarks.
\end{itemize}
The exchange of bosons allows different hadronic structures to be probed:
\begin{itemize}
\item
The prototype for {\bf existing hadronic structures} is the proton which
currently is the most precisely studied hadronic object.
\item
{\bf Genesis of hadronic structures} is analysed
using the structure developing in quantum fluctuations of the photon.
\item
{\bf Colour singlet exchange} constitutes a process beyond
single boson exchange. 
It's successful description provides a prime challenge for QCD.
\end{itemize}
It is the purpose of this contribution to underline these different
aspects of lepton-hadron scattering physics and their perspectives 
using as much as possible the measurements themselves.

\section{\boldmath Lepton--Quark Scattering at Attometer Distance
\label{sec:attometer}}

\begin{figure}[ttt]
\setlength{\unitlength}{1cm}
\begin{picture}(16.0,6.5)
\put(0.1,5.4)
{\epsfig{file=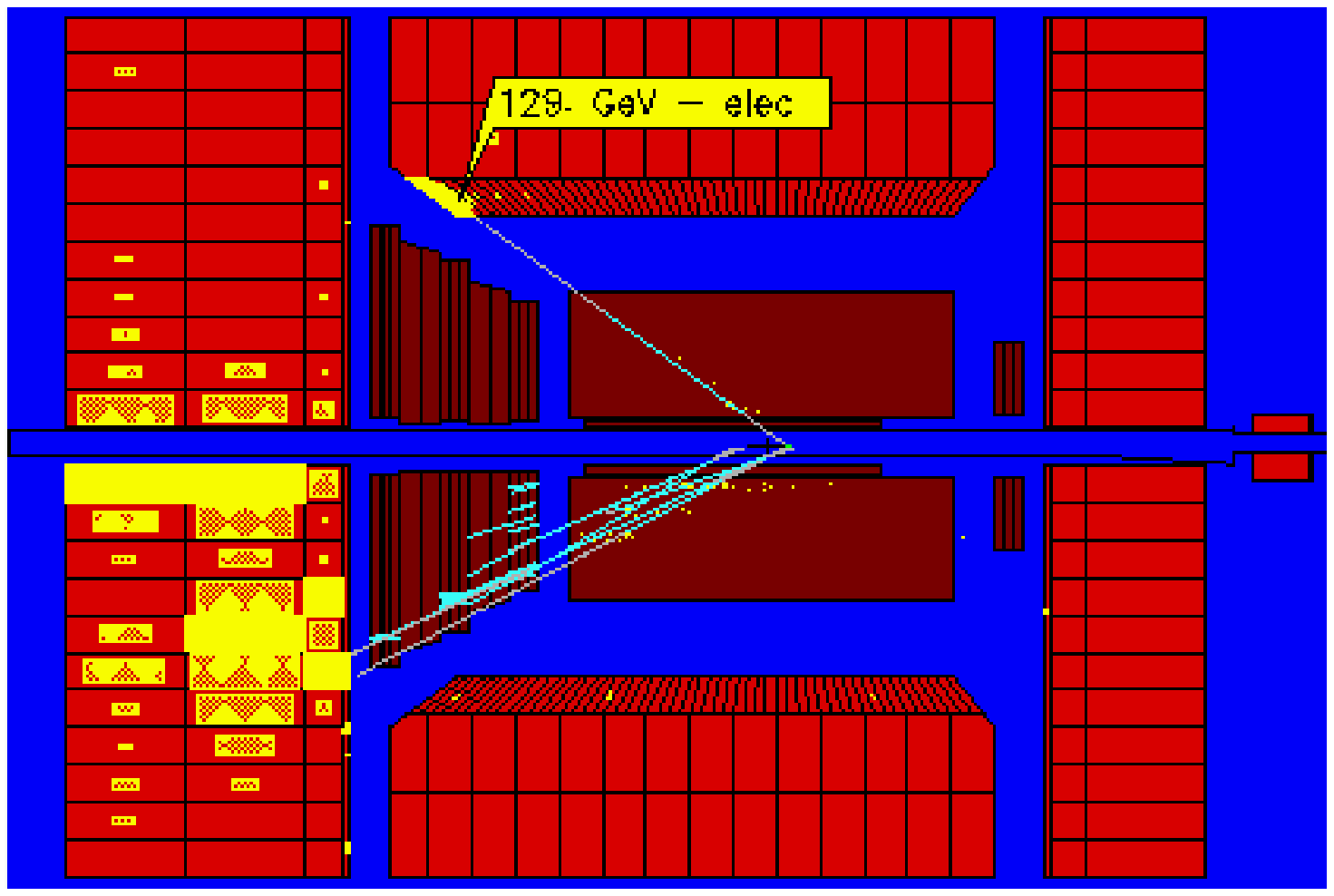,width=4.7cm,angle=270.}} 
\put(7.45,0.1)
{\epsfig{file=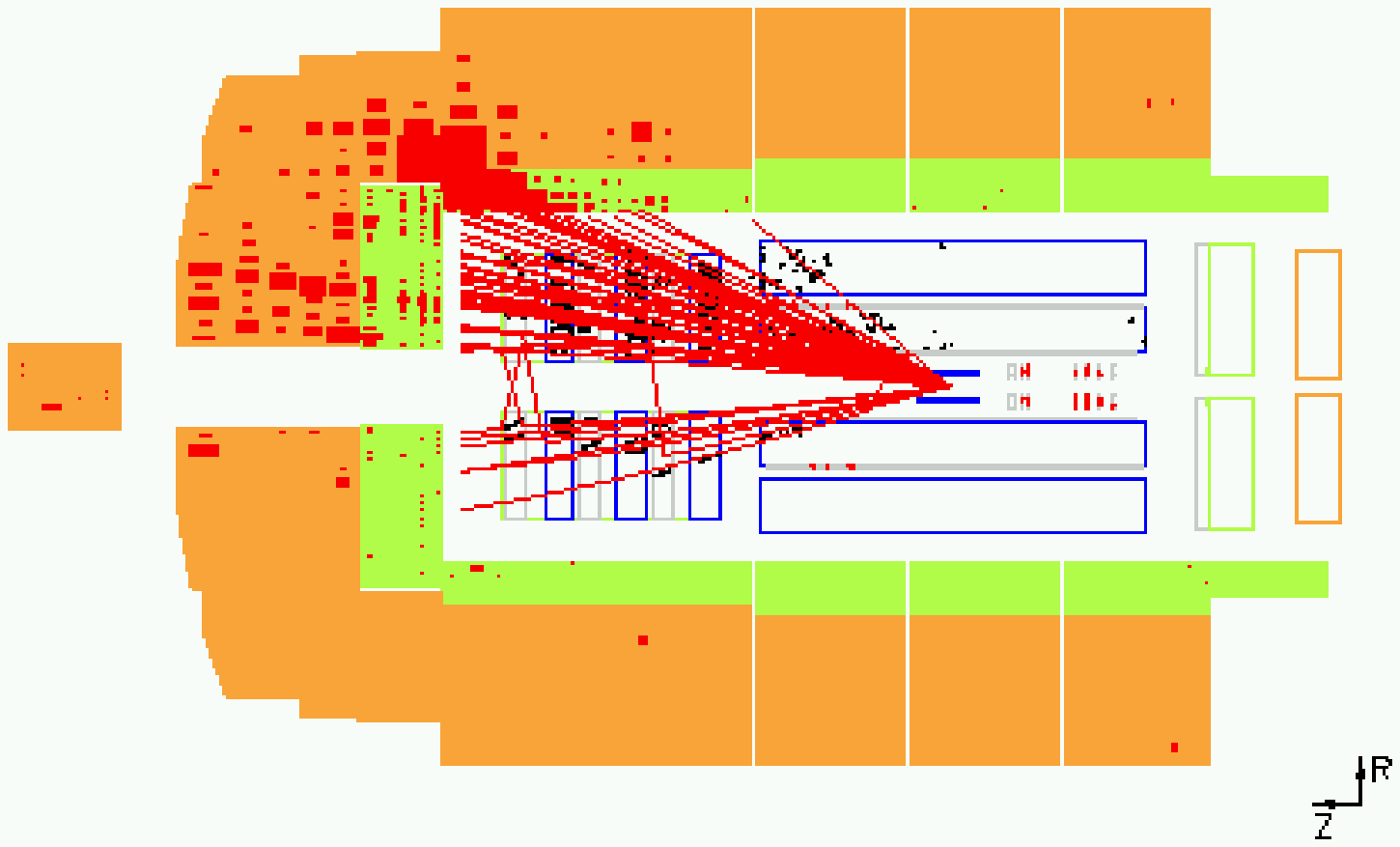,width=5.3cm,angle=90.0}}
\end{picture}
\caption{\label{fig:events}
Neutral and charged current interactions observed 
with the ZEUS and H1 experiments.}
\end{figure}

\noindent
The large center of mass energy at HERA of $\sqrt{s_{ep}}=318$ GeV 
allows lepton--quark scattering to be analysed at
distances down to almost \mbox{$1$ Attometer $=$} $10^{-18}$ m.
Both neutral and charged current interactions (Fig.\ref{fig:events})
are used to test the Standard Model predictions.
The double differential cross sections in terms of the 
resolution scale $Q^2$, which denotes the negative squared 
four-momentum transfer carried by the boson,
and the quark fractional momentum $x$ relative to the proton
are given by:
\begin{eqnarray}
\frac{{\rm d}^2\sigma_{NC}}{{\rm d}Q^2\, {\rm d}x} \; &\sim& \;
\alpha^2 \;\;\;\;\;\;\;\;\;\;\;\,\, \frac{1}{Q^4} \;\;\;\;\;\;\;\;\;\,
 \frac{1}{x} \; \Phi_{NC}( x, Q^2 ) \label{eq:nc}\\
\frac{{\rm d}^2\sigma_{CC}}{{\rm d}Q^2\, {\rm d}x} \; &\sim& \;
G_F^2 \; \left(\frac{M_W^2}{M_W^2+Q^2}\right)^2 \; \frac{1}{x} 
\; \Phi_{CC}\;( x, Q^2 ) \;\;\;. \label{eq:cc}
\end{eqnarray}
Here $\alpha$ and $G_F$ denote the coupling strength of the 
electromagnetic and weak interaction processes.
$M_W$ is the $W$-boson mass.
The $\Phi$ terms denote the spin 
characteristics of the scattering together with the probabilities
$x f(x,Q^2)$ of finding the different quark flavours in the proton.
In addition $\Phi_{NC}$ contains
terms for $Z$ exchange and $\gamma$-$Z$ interference.
At high $Q^2$ it is
\begin{eqnarray}
\Phi_{NC}^{e^\mp p} &\sim&
   \left( {1} + {\cos^4{\left( \frac{\theta^*}{2} \right) }} \right)
   \left( \frac{4}{9} ( {x u} + {x \bar{u}})
        + \frac{1}{9} ( {x d} + {x \bar{d}}\; ) \right) \;
\pm {\rm add. \, terms \, with} \, \gamma,Z \label{eq:phigz} \\
\Phi_{CC}^{e^-p} &\sim&
  {x u} \; + \;
{\cos^4{\left( \frac{\theta^*}{2}\right) }} \;\; {x \bar{d}}  \label{eq:phiw-}\\
\Phi_{CC}^{e^+p} &\sim&
  {x \bar{u}} \; + \;
{\cos^4{\left( \frac{\theta^*}{2} \right) }} \;\; {x d}  \label{eq:phiw+} \; . 
\end{eqnarray}
$\theta^*$ denotes the scattering angle in the lepton-quark center of mass system 
and can be calculated from 
$\cos^4{(\theta^*/2)} = ( 1 - Q^2/s_{ep}/x )^2$.
The two components in the angular distribution result from two spin 
configurations of the colliding lepton and quark:
if the spins add up to zero, any scattering angle is allowed. 
If the spins add up to $1$, backward scattering is forbidden 
for massless quarks and the angular 
distribution is weighted by $\cos^4{(\theta^*/2)}$. 
In charged current interactions also the quark type can be analysed:
e.g., positrons couple only to negatively charged quarks. 
In addition, right-handed positrons couple only to right-handed antiquarks,
or left-handed quarks.
This offers a unique handle to differentiate between 
quark flavours in the proton.

\begin{figure}[ttt]
\setlength{\unitlength}{1 cm}
\begin{center}
\begin{picture}(16.0,11.0)
\put(0,8){\Huge $\frac{d\sigma}{dQ^2}$}
\put(5.5,-0.2){\LARGE $Q^2$ [GeV$^2$]}
\put(3.5,10.6){\Large {Electron--Proton Collisions}}
\put(1.5,0.){\epsfig{file=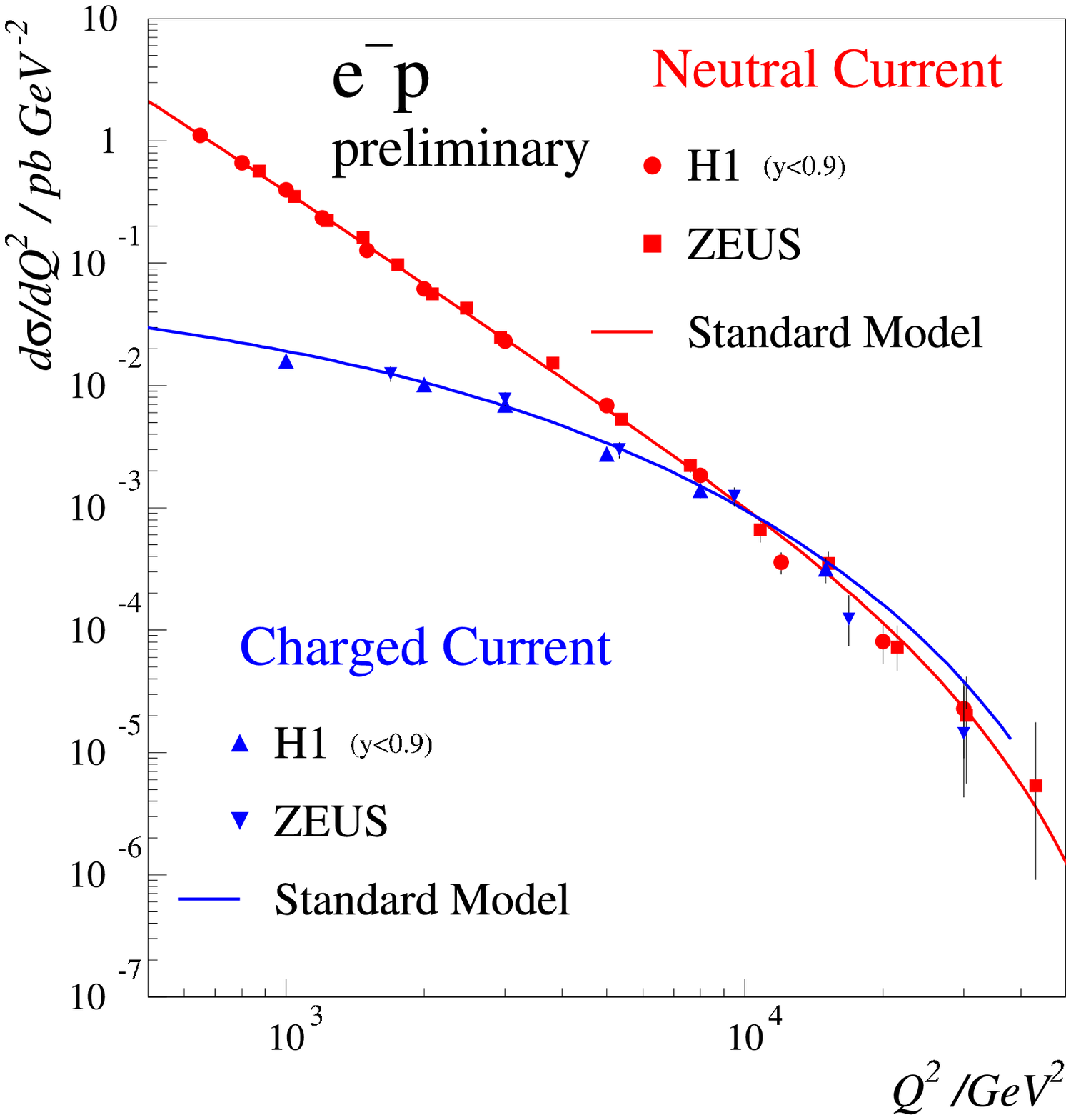,width=10cm}}
\put(13.,6.5){\epsfig{file=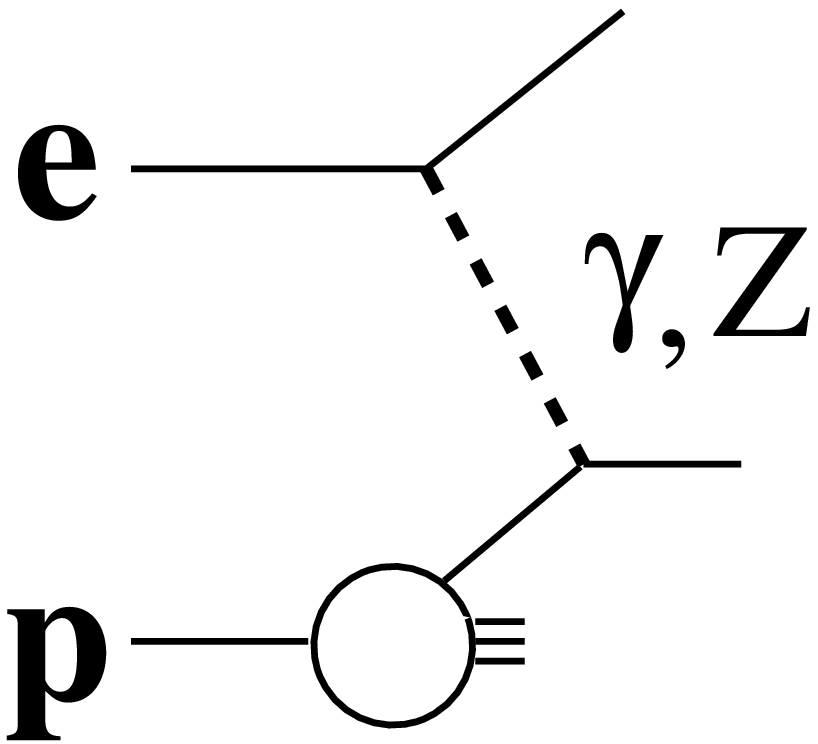,width=3cm}}
\put(13.,2.5){\epsfig{file=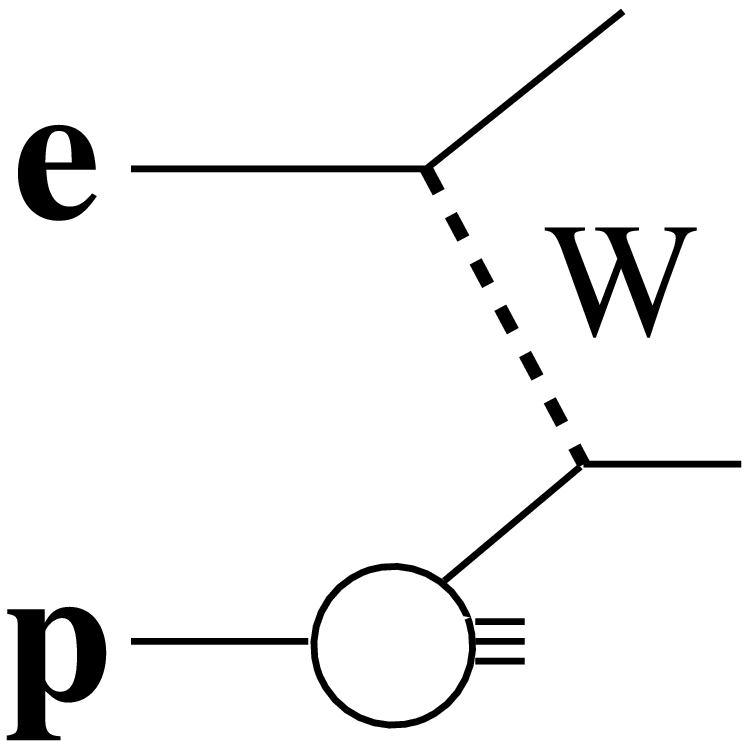,width=3cm}}
\end{picture}
\end{center}
\caption{\label{fig:xsecq2}
Differential cross section for neutral and charged current
interactions as a function of the resolution scale $Q^2$ from 
H1 and ZEUS data.}
\end{figure}

Integrating the double differential cross sections over $x$ gives the
single differential cross section which is
shown in Fig.\ref{fig:xsecq2} as a function of $Q^2$ from 
\mbox{H1 \cite{hih1}} and ZEUS \cite{hizeus} data.
Around $Q^2\sim 10^4$ GeV$^2$ 
the cross sections are found to be of equal magnitude.  
Since in both neutral current and charged current electron-proton scattering 
at high $Q^2$ primarily the $u$-valence quarks are probed,
eqs.~(\ref{eq:phigz}, \ref{eq:phiw-}), these data 
establish direct observation of the 
unification of the neutral current and charged current interactions at a
resolution scale corresponding to about \mbox{$10$ Attometer.} 

\begin{figure}[ttt]
\setlength{\unitlength}{1 cm}
\begin{center}
\begin{picture}(16.0,13.0)
\put(7.2,9.8){\Large {$xd$}}
\put(7.2,8.65){\Large {$x\bar{u}$}}
\put(0.6,8.99){---}
\put(7.0,8.99){---}
\put(3.2,-0.2)
{\LARGE $\cos^4{\frac{\theta^*}{2}}$ }
\put(0.0,0.5){\epsfig{file=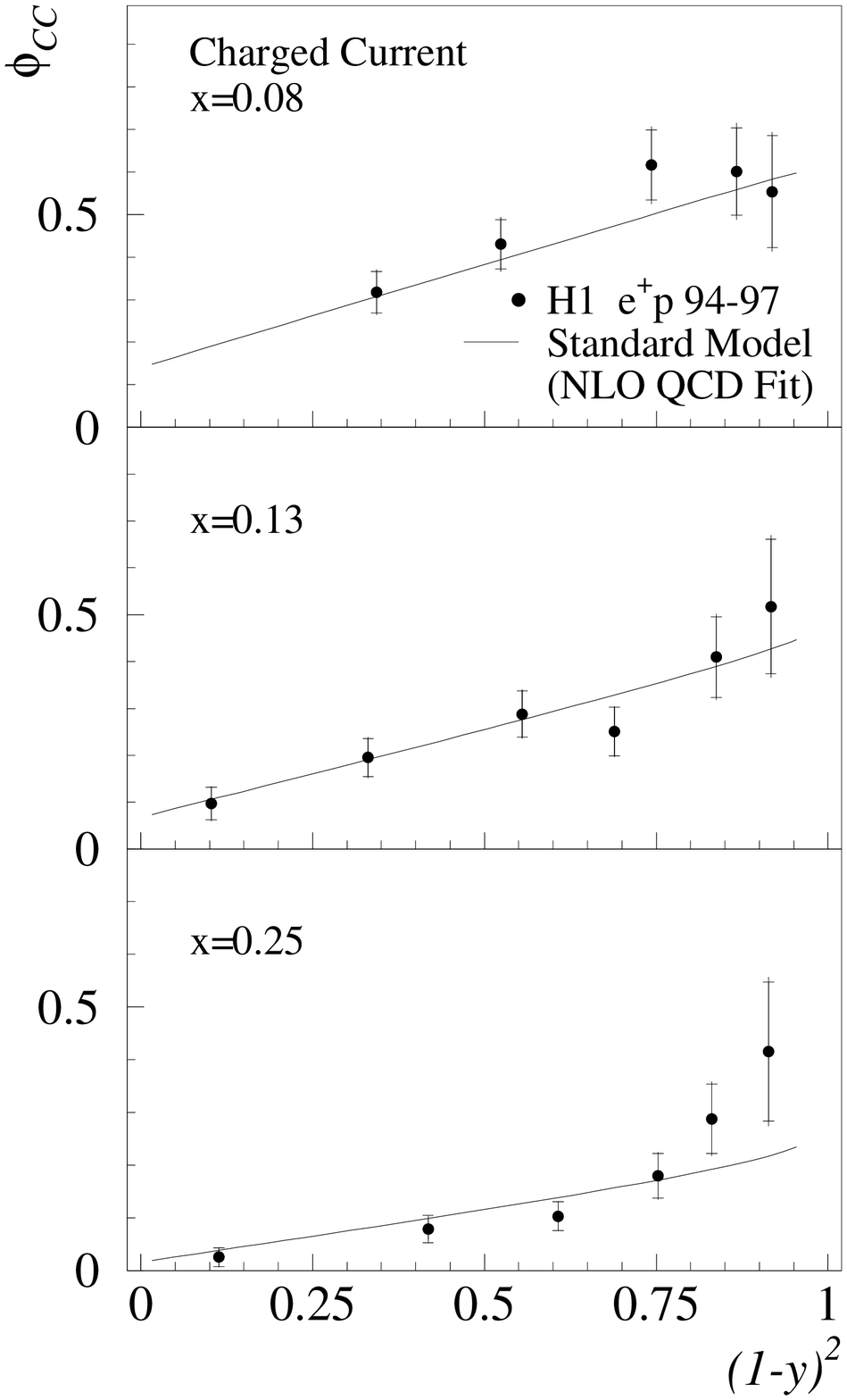,width=7.0cm}}
\put(15.2,10.7 ){\Large {$\frac{4}{9} xu$}}
\put(15.2,9.2 ){\Large {$\frac{4}{9} xu$}}
\put(8.7,10.02){---}
\put(15.1,10.02){---}
\put(11.2,-0.2)
{\LARGE $\cos^4{\frac{\theta^*}{2}}$ }
\put(8.1,0.5){\epsfig{file=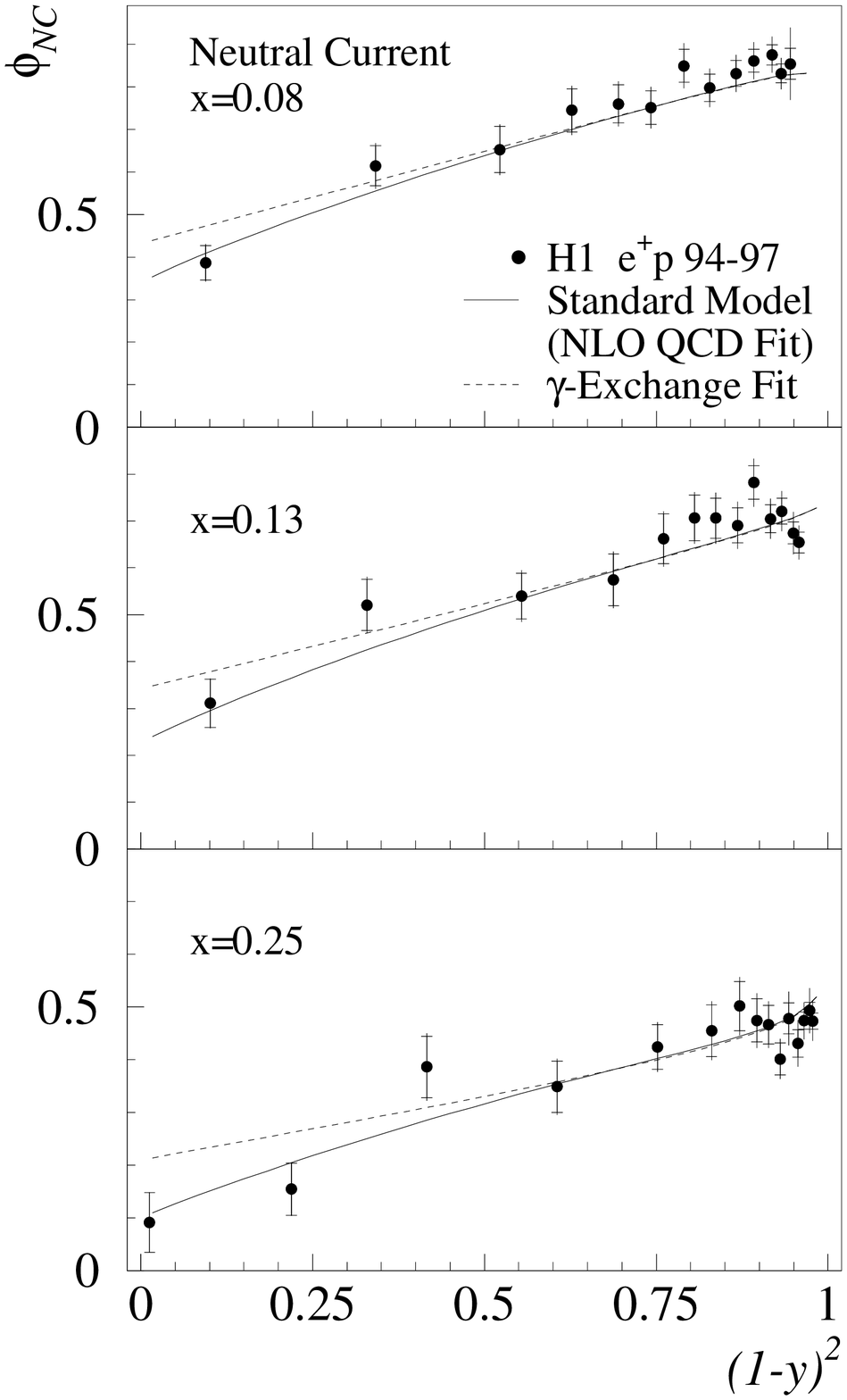,width=7.0cm}}
\end{picture}
\end{center}
\caption{\label{fig:spin} 
H1 measurements of the double differential charged and neutral 
current positron-proton
cross sections as a function of $\cos^4{(\theta^*/2)}$ 
in different bins of the parton fractional momentum $x$.
$\theta^*$ denotes the scattering angle in the 
lepton-quark center of mass system.}
\end{figure}

In Fig.\ref{fig:spin}, the spin characteristics of charged current 
positron-proton scattering is 
tested in the measurement of the weighted cross sections $\Phi$,
eq.~(\ref{eq:cc}), as a function of 
$\cos^4{(\theta^*/2)}$ \cite{h1hiq2}.  
Within the precision of the measurement, the data are in each $x$-bin compatible 
with a linear rise as expected from eq.~(\ref{eq:phiw+}). 
The extrapolation of the linear behavior to the backward scattering region 
($\cos^4{(\theta^*/2)}=0$) reveals a non-vanishing contribution of the 
negatively charged antiquarks, mainly $\bar{u}$.
Their relative contribution decreases as $x$ increases. 
The rising component reflects the contribution of the $d$-valence quarks. 
The $d$-quark density can be read off the forward scattering cross section 
($\cos^4{(\theta^*/2)}=1$).         

Also in neutral current interactions, the data are over a wide range 
compatible with a linear rise:
they reflect two equally large components, explained by the 
two spin configurations of the electromagnetic processes, 
eq.~(\ref{eq:phigz}). 
The forward scattering region ($\cos^4{(\theta^*/2)}=1$)
shows approximately the $u$-quark density in the proton:
$2 \times 4/9 \; xu \sim xu$.
In the region of backward scattering processes
($\cos^4{(\theta^*/2)}=0$), the cross section 
measurements deviate from the linear rise and demonstrate the onset 
of a new interaction:
the lower cross section results from the negative
interference between the photon and the $Z$-boson. 

The comparison of the positron-proton neutral current and charged 
current data in the forward scattering region of Fig.\ref{fig:spin}
demonstrates directly from the data that the $u$-quark density
is twice that of the $d$-quark.
Therefore the proton consists of the $uud$ quark configuration also at
the small distance scales probed at HERA.

The HERA luminosity upgrade program, starting to take data in 2001, 
is eagerly awaited:
much more precise data will challenge the Standard Model predictions 
for $ep$ processes in the Attometer regime.

\section{Existing Hadronic Structure: Proton \label{sec:proton}}

\noindent
As discussed in the previous section,
the $uud$ valence structure of the proton has been re-confirmed in the 
high $Q^2$ neutral and charged current measurements at HERA.
In the following our physics understanding of the 
proton structure function $F_2$ is discussed.
$F_2$ is determined from measurements of the double differential neutral
current cross section (compare with eqs.~(\ref{eq:nc}, \ref{eq:phigz}))
\begin{equation}
\frac{{\rm d}^2\sigma}{{\rm d}Q^2\, {\rm d}x} \; \sim \;
\alpha^2 \; \frac{1}{Q^4} \; \left( {1} + {\cos^4{\left( \frac{\theta^*}{2}\right) }} \right) \;
\frac{1}{x} \; F_2( x, Q^2 ) \label{eq:x}
\end{equation}
and contains the individual quark distributions, weighted by
the quark squared charges:
\begin{equation}
F_2(x, Q^2) \; \sim \; \frac{4}{9} ( xu  + x\bar{u} \;) + \frac{1}{9} ( xd + x\bar{d} \;) 
            + \frac{1}{9} ( xs + x\bar{s} \;) + ... 
\end{equation}

The QCD evolution equations predict that measurements of hadronic
structures depend on the logarithm of the resolution scale $Q^2$ at which
the structure is probed. 
On this basis, the following ansatz to analyse the $x$-dependence of
structure function data is explored \cite{martin}:
\begin{equation}
F_2(x,Q^2) = a(x) \; \left[ \ln{\left(\frac{Q^2}{\Lambda^2}\right)} 
\right]^{\;\; \kappa(x)} \; .
\label{eq:f2}
\end{equation}
Here $\Lambda$ is a scale parameter, $a$ reflects 
the charge squared weighted
quark distributions extrapolated to $\ln{(Q^2/\Lambda^2)}=1$, 
and $\kappa$ determines the positive and negative scaling violations of $F_2$.

\begin{figure}[ttt]
\setlength{\unitlength}{1cm}
\begin{picture}(12.0,15.0)
\put(2.5,7.5)
{\epsfig{file=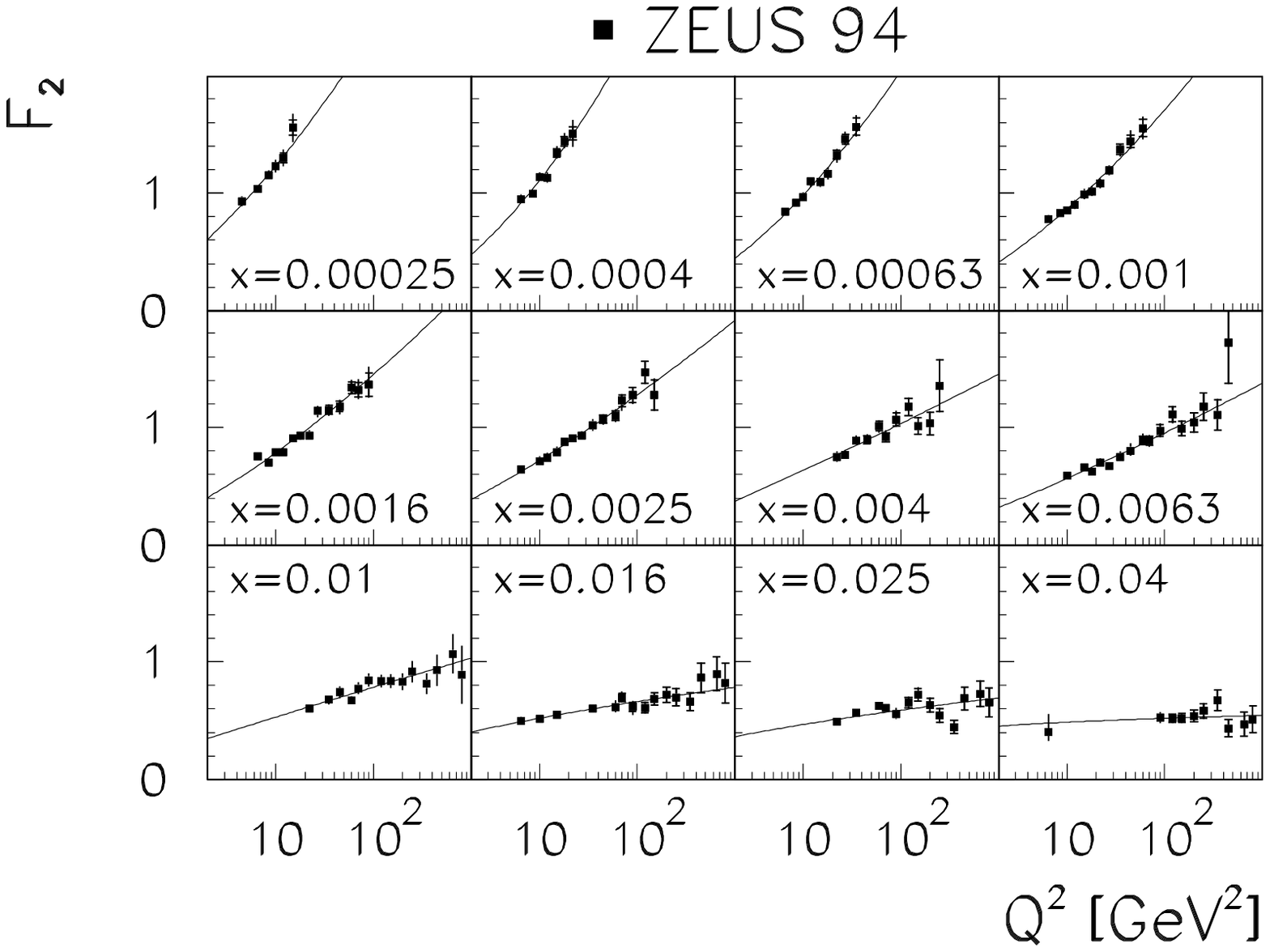,width=10.5cm}}
\put(2.5,0.0)
{\epsfig{file=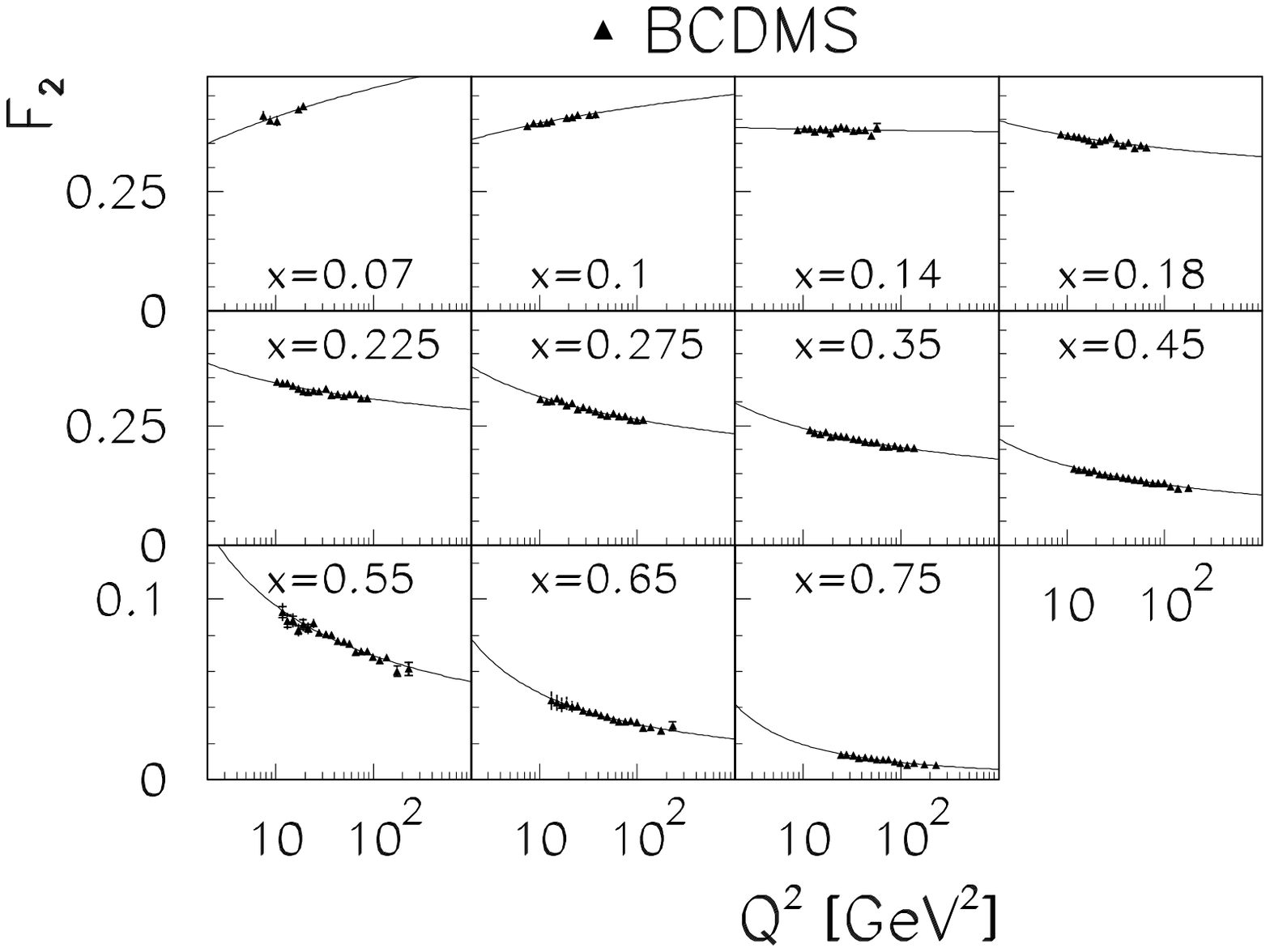,width=10.5cm}}
\end{picture}
\caption{\label{fig:proton}
ZEUS and BCDMS measurements of the proton structure function $F_2$
are shown as a function of $Q^2$ in the range $10^{-4} < x < 1$.
They are compared to the 2-parameter fits according to 
eq.~(\ref{eq:f2}) in each x-bin.}
\end{figure}

In Fig.\ref{fig:proton}, published ZEUS \cite{zeus94} low-$x$
data of the proton structure function $F_2$ for $Q^2>2$\, GeV$^2$ are shown.
In each $x$-bin, the result of a two-parameter fit according 
to eq.~(\ref{eq:f2})
is shown, using a fixed value of $\Lambda=0.35$ GeV
which represents a typical value of the strong interaction scale.
Only the total experimental errors have been used, ignoring 
correlations between individual data points.
The same fitting procedure has been applied to BCDMS data \cite{bcdms}
which are taken here as a reference sample for the high-$x$ region.

\begin{figure}[ttt]
\setlength{\unitlength}{1cm}
\begin{picture}(12.0,14.0)
\put(4.0,13.6)
{{\Large sea quarks}}
\put(9.6,13.6)
{{\Large valence quarks}}
\put(1.5,-0.5)
{\epsfig{file=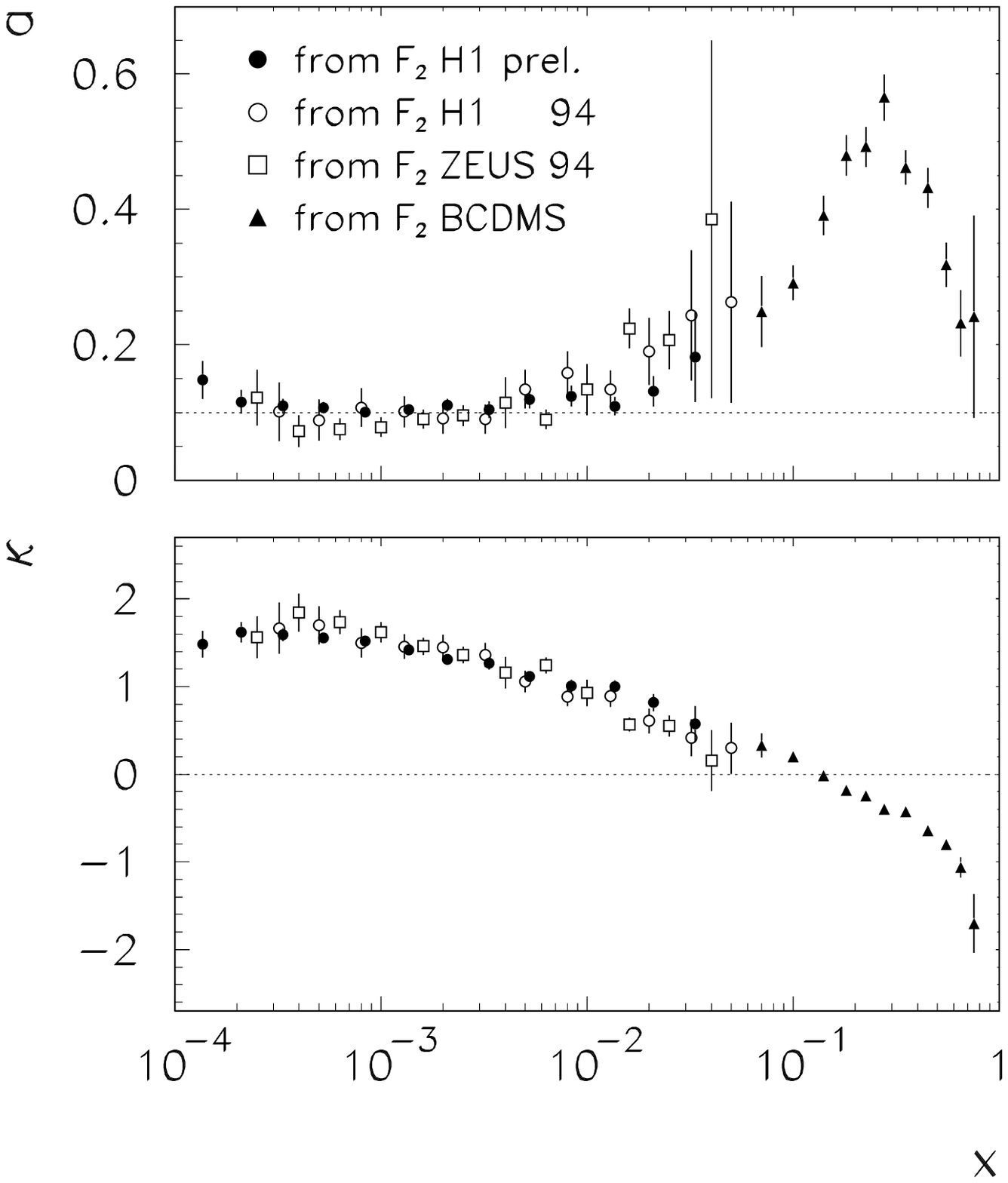,width=13cm}}
\end{picture}
\caption{\label{fig:akappa} 
The quark distribution $a(x)$ of the proton extrapolated to $Q^2 = 0.3$ GeV$^2$
and the scaling violations $\kappa(x)$ from the fits to the published
H1, ZEUS, BCDMS, and to the H1 preliminary data according to 
eq.~(\ref{eq:f2}).
The dotted lines serve to guide the eye.}
\end{figure}

The resulting parameters $a$ and $\kappa$ are summarized in 
Fig.\ref{fig:akappa} as a function of $x$ together with fits
to the published H1 low-$x$ data \cite{h194}. 
Also shown are fits to the preliminary H1 data \cite{h1prel}
which are much more precise than the previous measurements.

For $a$, the data fits exhibit two distinct regions: 
around $x\sim 0.3$ they reflect the valence quark distributions,
implying that each valence quark carries $\sim 1/3$ of the proton 
momentum.
At low $x$, $a(x)$ is
compatible with converging to a constant value.
A comparison of the lowest point, derived from the H1 preliminary 
measurement, with the new ZEUS preliminary data presented at this conference 
will be of interest. 
The resulting scaling violation term $\kappa$ appears to rise as $x$ 
decreases, exhibiting the negative and positive scaling violations of $F_2$
for $x$ above and below $0.1$ respectively.
The errors in Fig.\ref{fig:akappa} represent the statistical errors of 
the fits.
Both parameters $a$ and $\kappa$ are anti-correlated as can be seen from 
neighbouring points.
No significant $Q^2$-dependence of $a$ and $\kappa$
has been found in the published data 
when the fits were repeated for two intervals in $Q^2$ 
(above and below $20$ GeV$^2$).

With $a$ being approximately constant below $x\sim 10^{-2}$, 
changes of $F_2$ at low $x$ result from 
the scaling violation term $\kappa$ alone, indicative of 
the interaction dynamics that drives $F_2$
and in support of the predictions \cite{lowx,grv}.

The parameter $a$ has already been identified as the 
charge squared weighted quark distributions
extrapolated to $\ln{(Q^2/\Lambda^2)}=1$ which corresponds
here to $Q^2 = 0.3$ GeV$^2$.
An understanding of the parameters $\Lambda$ and $\kappa$ can be
achieved by comparison with the QCD evolution equation which is 
written here in the leading order DGLAP approximation:
\begin{equation}
\frac{{\rm d} f_i(x,Q^2)}{{\rm d} \ln{Q^2}} =
\frac{\alpha_s(Q^2)}{2\pi} \sum_j \int_x^1 \frac{{\rm d}y}{y} 
P_{ij}\left( \frac{x}{y} \right) f_j(y,Q^2)  \; .
\label{eq:dglap}
\end{equation}
Here $f_i, f_j$ denote the parton densities, $P_{ij}$
are the splitting functions, and
\begin{equation}
\alpha_s = \frac{b}{\ln{(Q^2/\Lambda_{QCD}^2)}}
\label{eq:alphas}
\end{equation}
is the strong coupling constant.

The derivative of the ansatz chosen here, eq.~(\ref{eq:f2}),
with respect to $\ln{Q^2}$ gives
\begin{equation}
\frac{{\rm d} F_2(x,Q^2)}{{\rm d} \ln{Q^2}} 
= \frac{1}{\ln{(Q^2/\Lambda^2)}} \; \kappa(x)  \; F_2(x,Q^2) \; ,
\label{eq:f2t}
\end{equation}
where relating $1/\ln{(Q^2/\Lambda^2)}$ 
with $\alpha_s$ in eq.~(\ref{eq:dglap}) 
implies association of the scale parameter $\Lambda$ in eq.~(\ref{eq:f2}) 
with the QCD parameter $\Lambda_{QCD}$.
The term $\kappa$ relates to the sum over the different 
parton radiation terms in eq.~(\ref{eq:dglap})
divided by $F_2$.
$\kappa$ increases towards small $x$, consistent with larger phase space 
available for parton radiation.

To match the description chosen here, eq.~(\ref{eq:f2}), 
with the double asymptotic approximation expected from QCD for the 
gluon-dominated region at small $x$, 
$F_2\sim \exp{\sqrt{-\ln{x} \; \ln{(\ln{(Q^2/\Lambda^2)})}}}$
\cite{lowx,das},
the scaling violation term $\kappa$ is required to have a dependence like 
\begin{equation}
\kappa\sim \sqrt{\frac{-\ln{x}}{\ln{(\ln{(Q^2/\Lambda^2)})}}} \;\; .
\label{eq:das}
\end{equation}
The $Q^2$-dependence of $\kappa$ is therefore expected to be very small
which is in agreement with the experimental observation stated above.

More precise data and data reaching smaller values of $x$ will 
determine whether or not the scaling violations further increase 
towards low-$x$ and therefore give valuable information on
the parton densities in the proton as $x$ approaches $0$.

\section{Genesis of Hadronic Structure: Photon \label{sec:photon}}

\begin{figure}[ttt]
\setlength{\unitlength}{1cm}
\begin{picture}(12.0,7.0)
\put(12.4,2)
{\epsfig{file=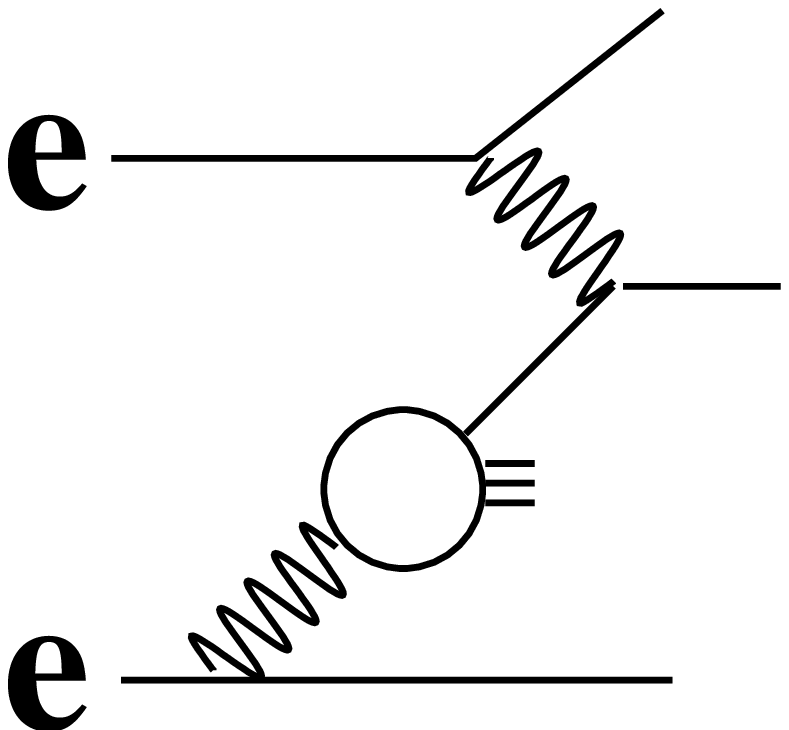,width=3cm}}
\put(0.,-0.5)
{\epsfig{file=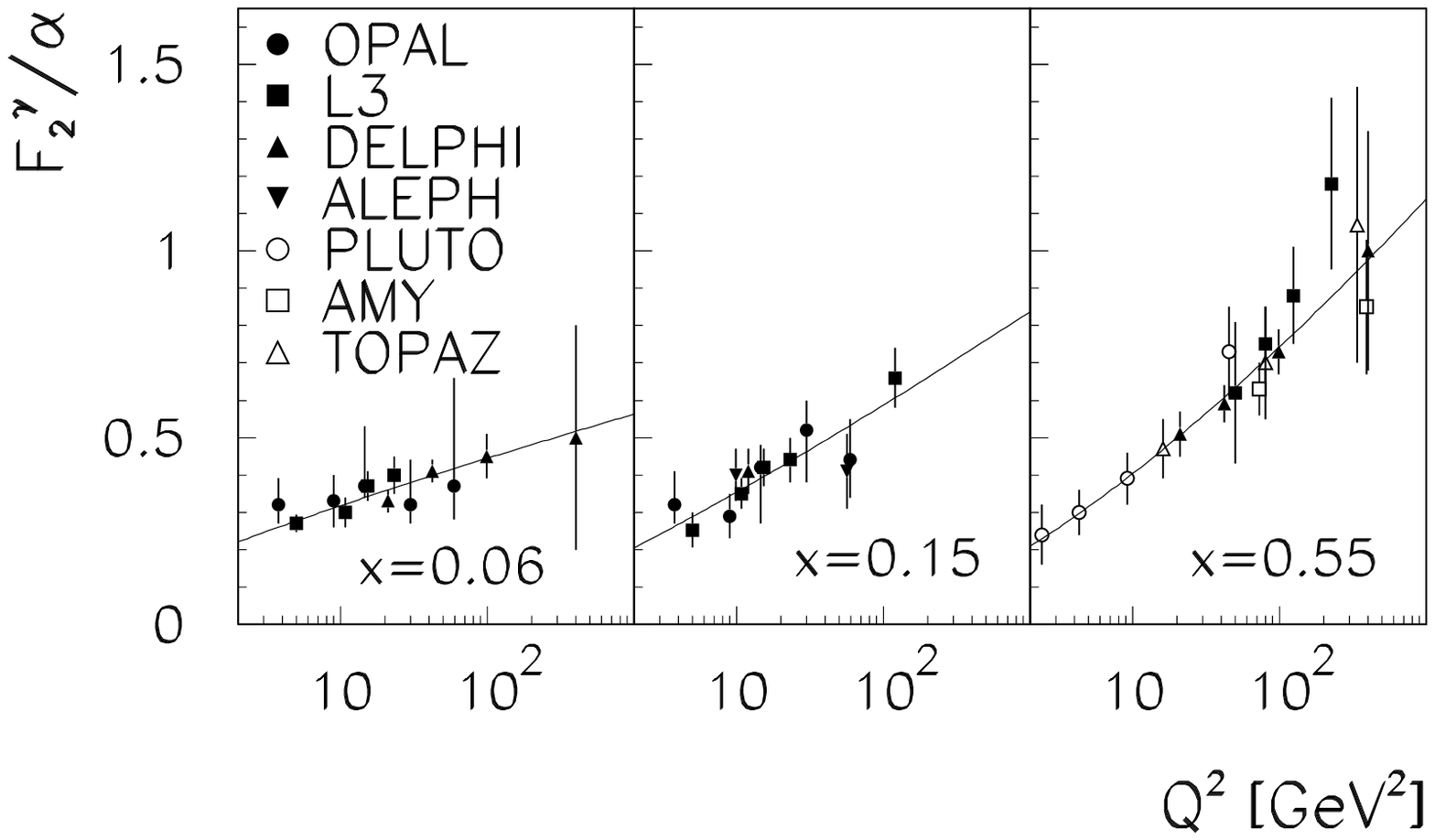,width=12cm}}
\end{picture}
\caption{\label{fig:photon} 
Measurements of the photon structure function are shown as a function
of $Q^2$.
They are compared
to the 2-parameter fits according to eq.~(\ref{eq:f2}) in each $x$-bin.}
\end{figure}

\noindent
The photon structure results from fluctuations of a photon
into a colour neutral and flavour neutral hadronic state.
For comparison with the proton data, the same fits according to
eq.~(\ref{eq:f2}) have been applied to recent measurements of the
photon structure function $F_2^\gamma$ which have been performed at
$e^+e^-$ colliders \cite{richard} (Fig.\ref{fig:photon}).

\begin{figure}[ttt]
\setlength{\unitlength}{1cm}
\begin{picture}(12.0,14.0)
\put(1.5,-0.5)
{\epsfig{file=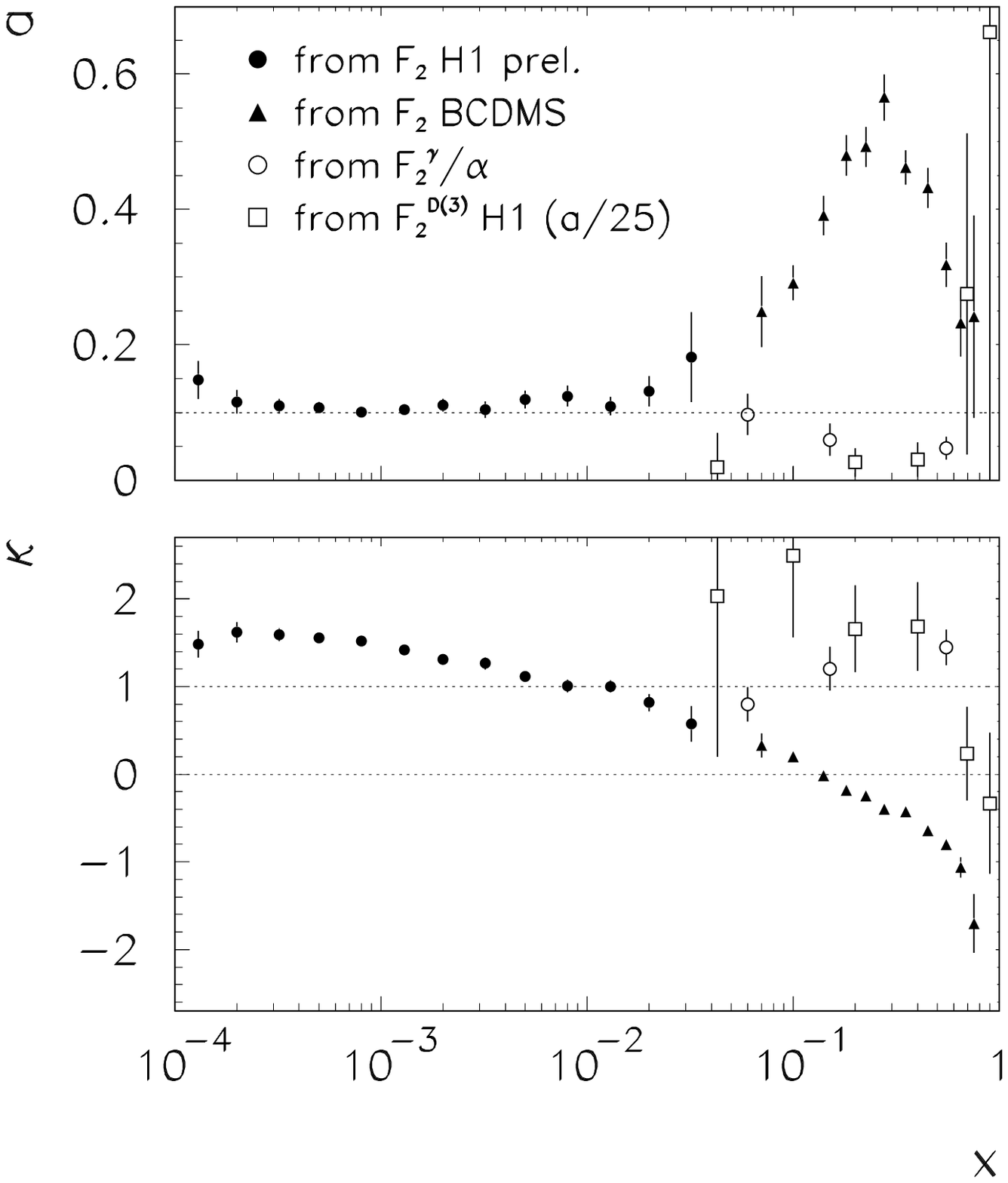,width=13cm}}
\put(5.0,6.45)
{\epsfig{file=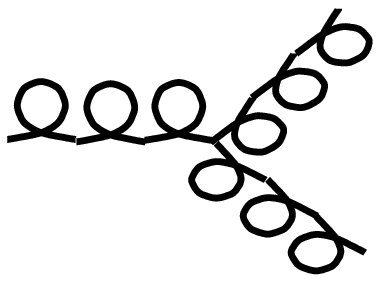,width=1cm}}
\put(11.25,6.45)
{\epsfig{file=gluon-split.eps,width=1cm}}
\put(11.25,5.1)
{\epsfig{file=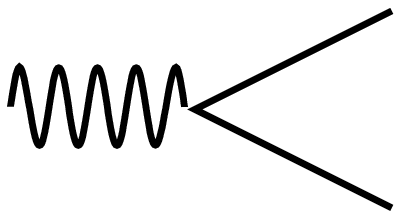,width=1cm}}
\put(11.25,3.25)
{\epsfig{file=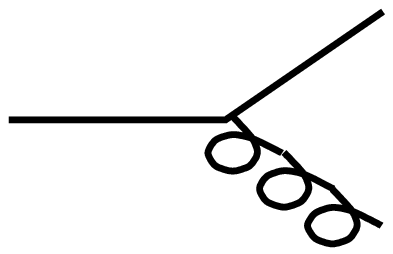,width=1cm}}
\end{picture}
\caption{\label{fig:comparison} 
The hadronic structures $a(x)$, extrapolated to $Q^2=0.3$ GeV$^2$,
and the scaling violations $\kappa(x)$ from fits to structure function 
data according to eq.~(\ref{eq:f2}) are compared between 
the proton, photon, and colour singlet exchange.
The diagrams of splitting functions indicate regions they contribute
to the QCD evolution.
The lines serve to guide the eye.}
\end{figure}

The values of the parameters $a$ and $\kappa$ are shown 
in Fig.\ref{fig:comparison} as the open circles.
Both parameters are distinct from those of a hadronic bound state 
like the proton:
in $a(x)$ the photon data exhibit no valence quark structure.
Instead, in the low-$x$ region around 
$x\sim 0.1$ the photon data prefer similar values of $a$ to the 
proton data for $x\le 0.01$.
The scaling violations $\kappa$ are positive at all values of $x$
and $\kappa$ is approximately $1$.
This is as expected from QCD calculations which predict
$F_2^\gamma$ for $0.1 < x < 1$ \cite{witten}.

Judgement on a universal low-$x$ behaviour of hadronic structures
will result from more precise measurements and lower-$x$ data 
of the photon structure function.
If the photon data show a constant quark density at small $x$ 
similar to the low-$x$ proton data, scaling violations of $F_2^\gamma$,
which deviate from those resulting from the photon splitting into 
quark-antiquark pairs and approach those observed 
for the proton, could become visible below or slightly above $x=10^{-2}$
where also for the proton data it is $\kappa \sim 1$.

\begin{figure}[ttt]
\setlength{\unitlength}{1cm}
\begin{picture}(12.0,10)
\put(0.,-0.5)
{\epsfig{file=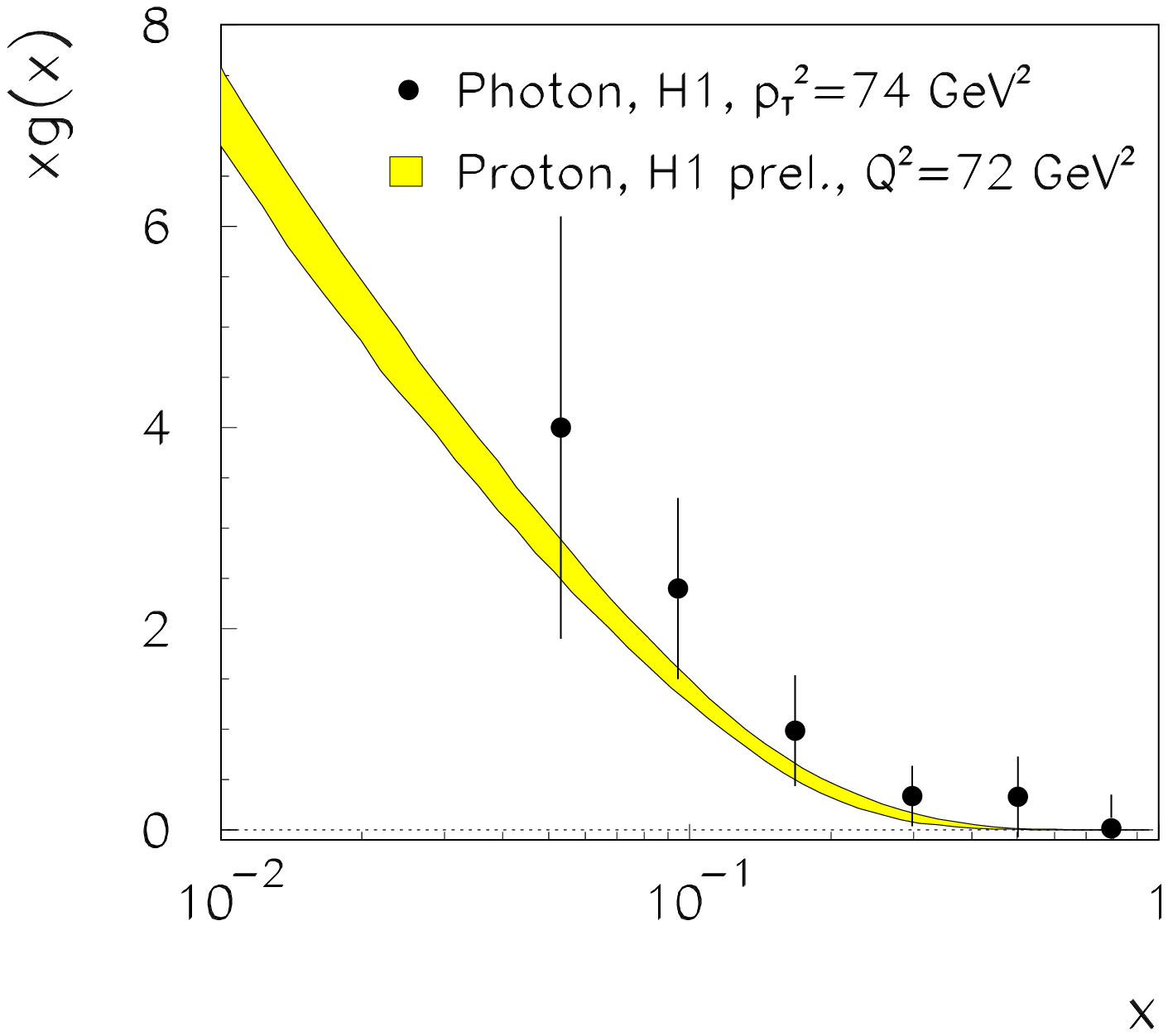,width=12cm}}
\put(12.5,6.5)
{\epsfig{file=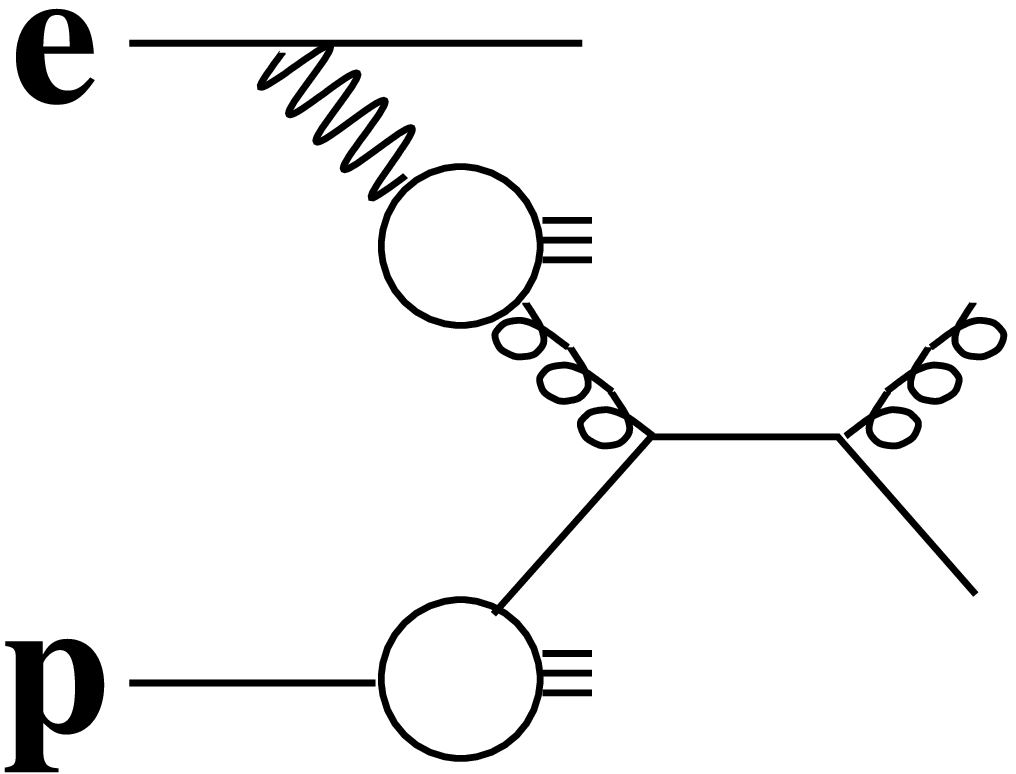,width=3cm}}
\put(12.5,2)
{\epsfig{file=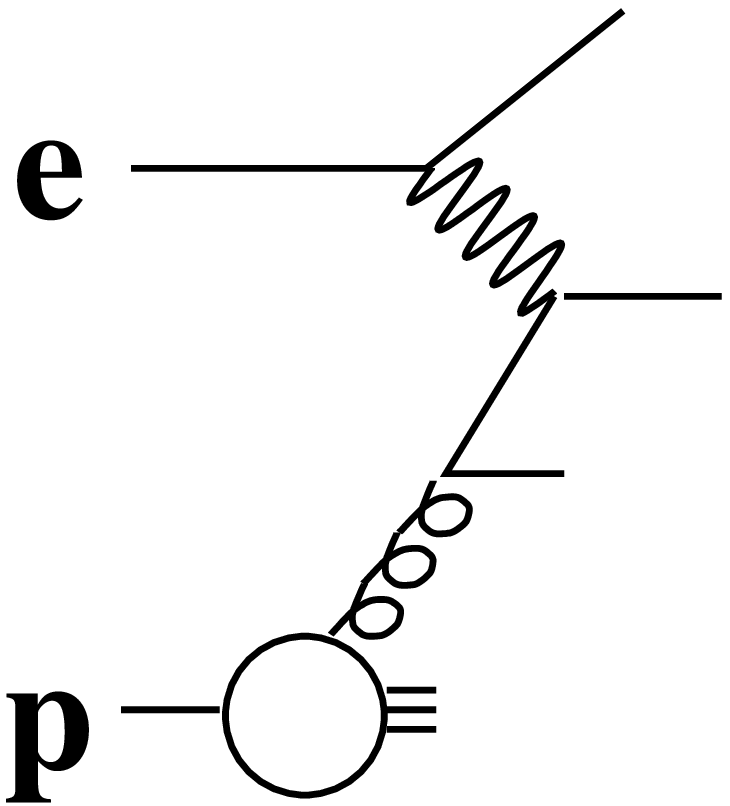,width=2.5cm}}
\end{picture}
\caption{\label{fig:gluon} 
Comparison of the H1 photon and proton gluon distributions as a function
of $x$.}
\end{figure}

Interesting information on the question of universality comes already from measurements of
the gluon in the photon probed in strong interaction 
processes in photon-proton collisions. 
The production of two-jet events is sensitive to the gluons developing in  
photon fluctuations. 
In Fig.\ref{fig:gluon}, a recent measurement of $x g(x)$ is shown \cite{h1gluon}. 
The gluons appear as the low-$x$ companions of 
the newly built hadronic structure:
at large $x$ the gluon density is small;
it rises towards small values of $x$.

In the same figure, this gluon distribution is compared to the gluon distribution 
of the proton, determined from measurements of the proton structure function
\cite{h1prel}.
Although the error bars of the photon measurement are large and $Q^2$ and 
$p_t^2$ may not represent the very same resolution scale, the similarity of the 
newly built and the already existing gluon distribution is striking. 
This observation may be a first experimental indication of a universal gluon 
distribution developing in hadronic structures.                   

\section{Colour Singlet Exchange \label{sec:cse}}

\begin{figure}[ttt]
\setlength{\unitlength}{1cm}
\begin{picture}(12.0,8)
\put(0.,-0.5)
{\epsfig{file=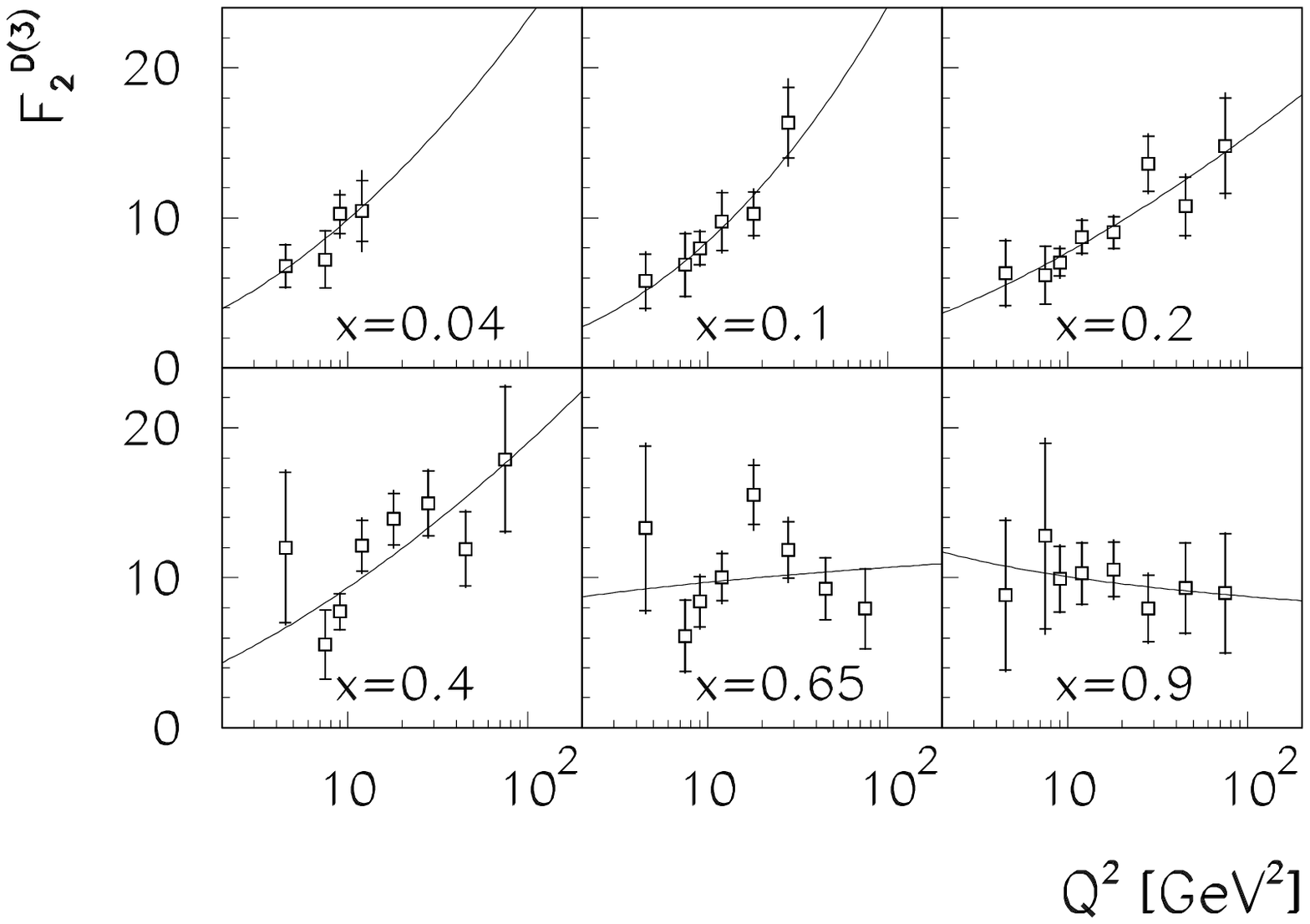,width=12cm}}
\put(12.7,2.6)
{\epsfig{file=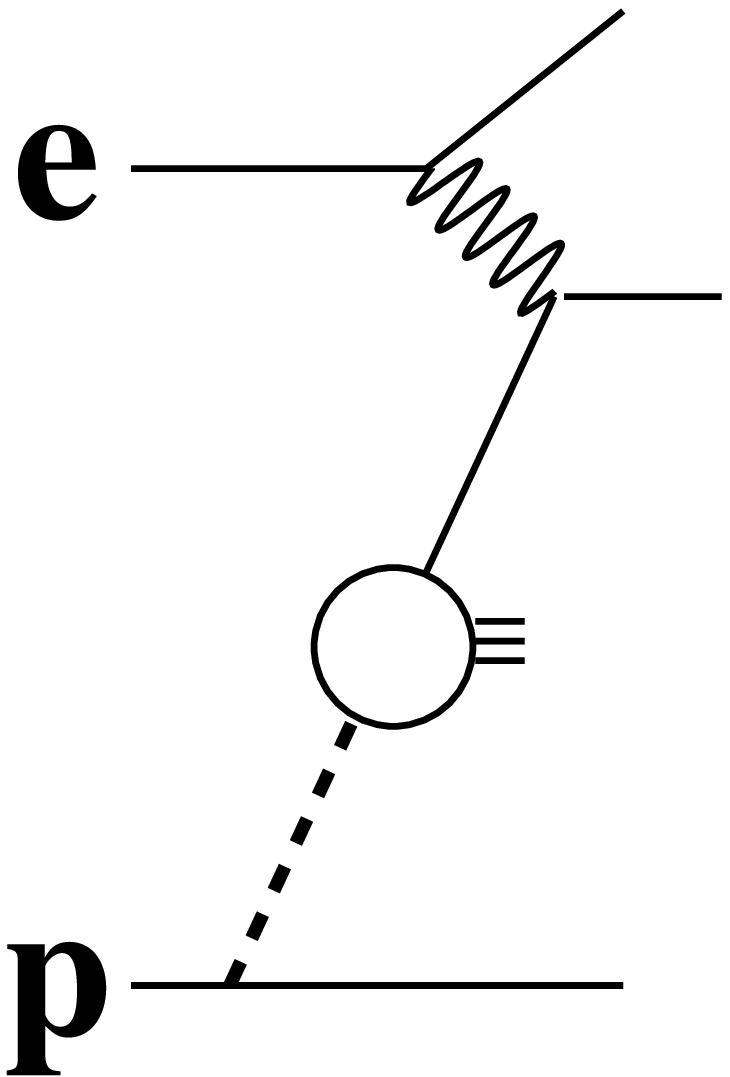,width=2.6cm}}
\end{picture}
\caption{\label{fig:cse} 
H1 measurements of the structure function of colour singlet exchange
are shown as a function of $Q^2$.
They are compared
to the 2-parameter fits according to eq.~(\ref{eq:f2}) in each $x$-bin.}
\end{figure}

\noindent
Further information on gluons in hadronic structures results from 
structure function measurements of
colour singlet exchange.
In Fig.\ref{fig:cse}, H1 $F_2^{D(3)}$ \mbox{data \cite{diff}}
are compared to the same two-parameter 
fits as used above, eq.~(\ref{eq:f2}).
Here $x$ (frequently called $\beta$) 
denotes the fractional momentum of the scattered parton
relative to the colour neutral object, which itself carries a
fractional momentum $\xi=0.003$ relative to the proton
and therefore belongs to the low-$x$ companions of the proton.

Also these data exhibit scaling violations $\kappa$ that
are different from the proton measurements at the same values of $x$
(Fig.\ref{fig:comparison}). 
Instead, for $x<0.5$ they have the tendency of being larger 
than the photon data and are similar to the low-$x$ proton data.
The large rate of events with colour singlet exchange together with the
large scaling violations of $F_2^{D(3)}$ is suggestive 
of a gluon dominated exchange.

The values of the normalization $a$ rise towards $x=1$ 
to about $a=10$ (in Fig.~\ref{fig:comparison}, the parameter $a$
has been scaled by $1/25$).
These values have large uncertainties of the order of $100\%$.
If more precise data support such a singular parton density 
for $x\rightarrow 1$ at low $Q^2$, 
then these colour neutral fluctuations consist of one
gluon carrying 
essentially all the colour singlet momentum and (at least) one 
further gluon with very low momentum neutralizing the colour.

\section{Predictive Power for Proton Interaction Processes}

\noindent
The proton structure reveals amazing simplicity:
at low resolution scale $Q^2$, the three valence quarks $uud$
each carry fractional momentum $x = 1/3$
(see sections \ref{sec:attometer}, \ref{sec:proton}).
The sea quark contribution is at low values of $x$ independent of $x$
(see section \ref{sec:proton}).
Gluons accompany the proton at low $x$ with a 
possibly universal momentum distribution (see section \ref{sec:photon}).
Gluons initiate colour neutral configurations
together with other very low momentum gluons (see section \ref{sec:cse}).

\begin{figure}[ttt]
\setlength{\unitlength}{1 cm}
\begin{center}
\begin{picture}(16.0,7.5)
\put(7.0,-0.5)
{\LARGE $x$}
\put(15.2,-0.5)
{\LARGE $x$}
\put(0.0,6.0)
{\huge $xq_i$}
\put(8.9,5.8)
{\Huge $\frac{d}{u}$}
\put(1.2,-0.1)
{\epsfig{file=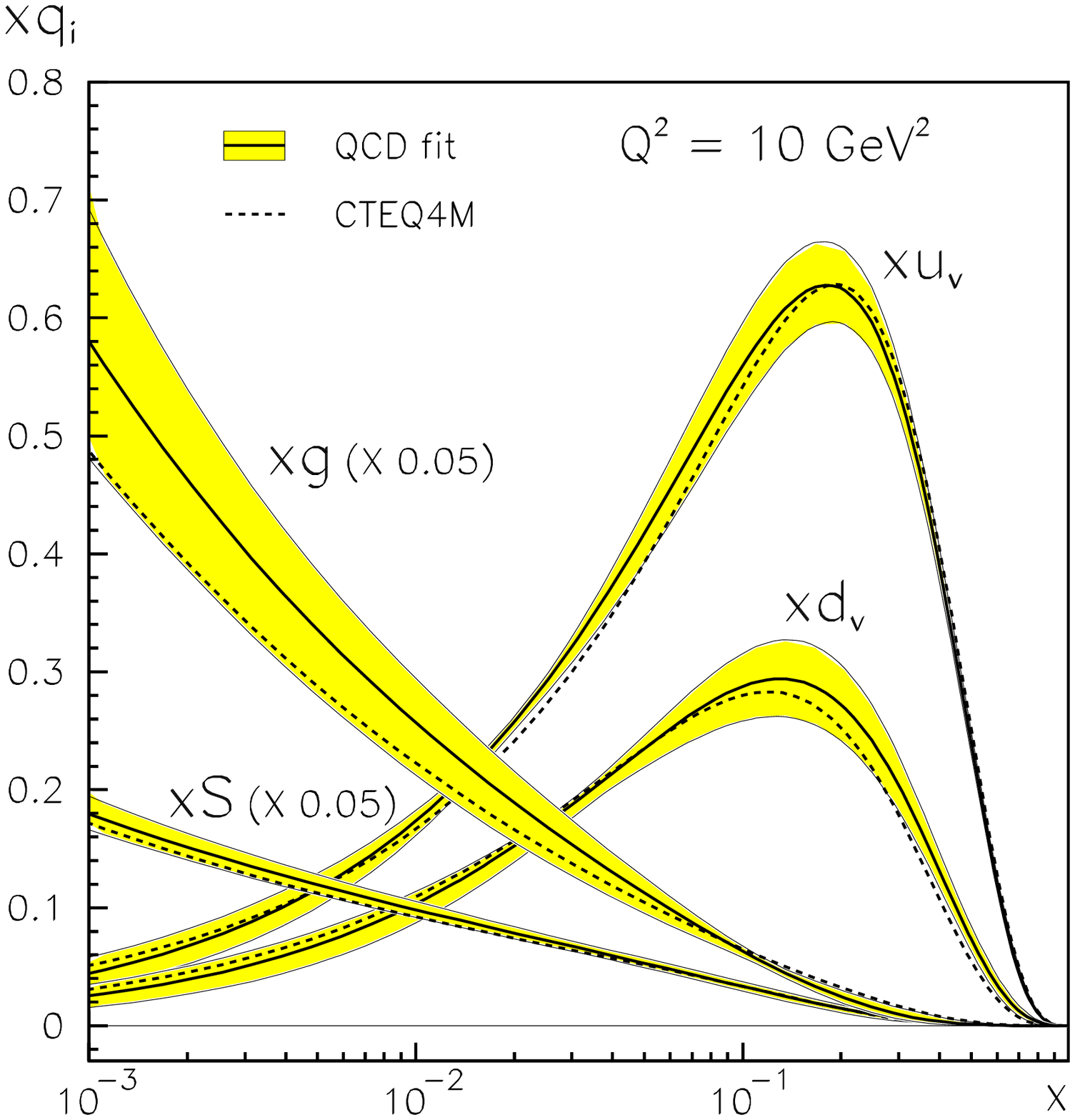,width=7cm}}
\put(9.6,0.0)
{\epsfig{file=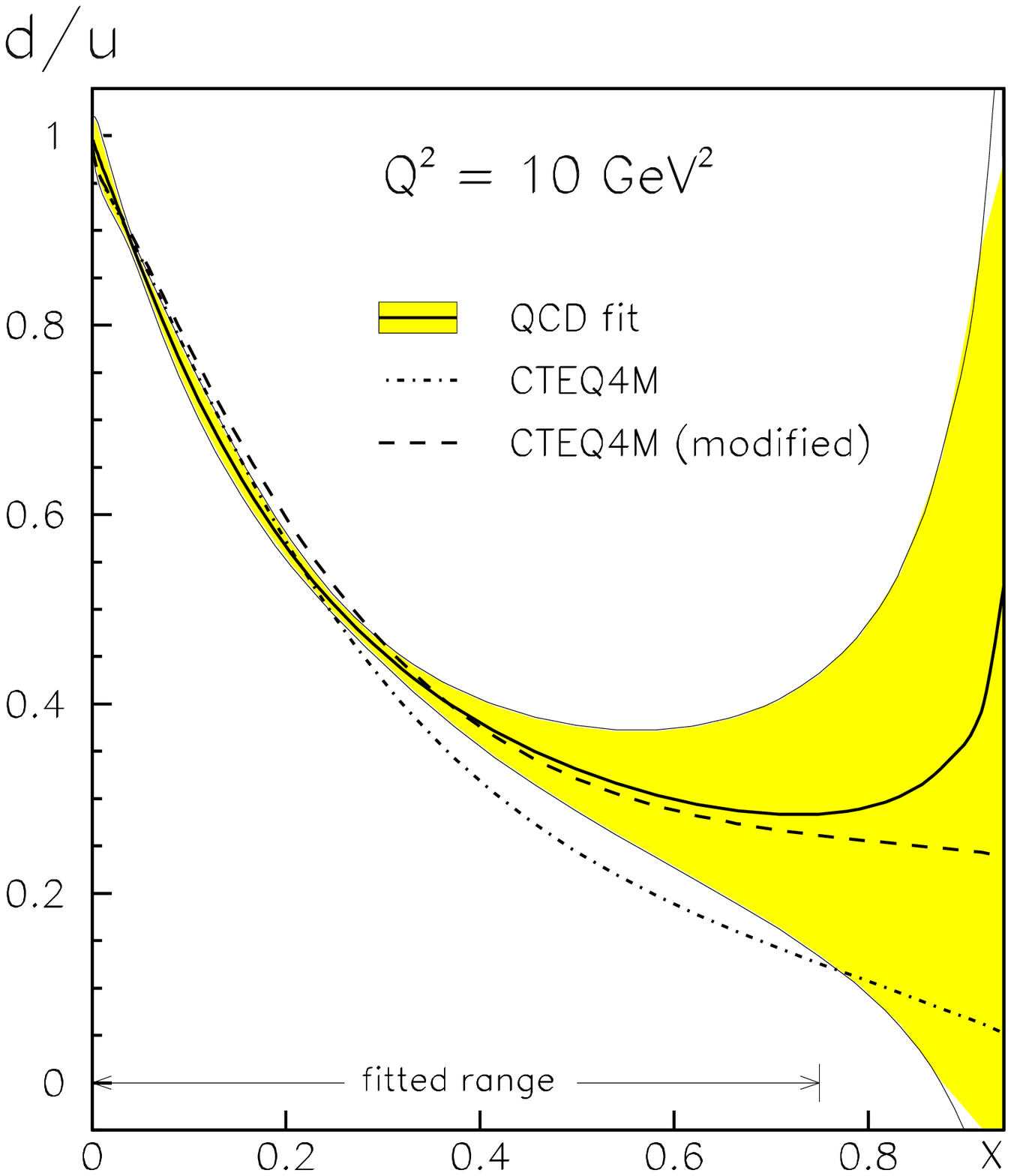,width=6.3cm}}
\end{picture}
\end{center}
\caption{\label{fig:botje} 
Parton distribution functions and $d/u$ ratio as a function of $x$
at $Q^2=10$ GeV$^2$ from a global fit which takes into account
experimental errors and correlations between the individual parton
distribution functions.}
\end{figure}

However, to predict interactions with protons, 
full information on all individual parton distributions of the
proton are required.
While such parton distribution functions $xf_i$ have been available from 
global fits for many years, recent pioneering work has
succeeded in determining the precision of these distribution
functions taking into account the precision of the measurements
and correlations between the different functions $xf_i$ \cite{botje}
(Fig.\ref{fig:botje}).
This analysis shows a good knowledge of the functions $xf_i$ over 
a wide range in $x$.
However, the knowledge for $x\rightarrow 1$ is not satisfactory:
large values of $x$ correspond to high resolution power
at hadron colliders (e.g. LHC)
and point at the potential discovery region for new physics.
An improved determination of the proton parton distribution 
as $x$ approaches $1$ by deep inelastic scattering experiments
is therefore mandatory and currently is under discussion \cite{temple}.

Further questions on the predictive power of QCD calculations
for proton-proton interactions result from the measurements
of forward jet and forward $\pi^\circ$ cross sections in $ep$ collisions,
e.g. \cite{wengler}.
These measurements explicitely test QCD evolution over some rapidity
distance and may signal limitations of the current approximations of
QCD evolution to simple process configurations at small distances.
Here theoretical work is needed and ongoing.

\section{Achievements and Challenges}

\noindent
We currently celebrate the $30$ years knowledge of valence quarks in 
the proton.
The new contribution of the HERA collider experiments
to the understanding of the proton is the
low-$x$ structure which appears as a consequence of QCD dynamics.
Open questions are:
is the parton density of the proton finite as $x\rightarrow 0$ ?
What is the parton density at $x\rightarrow 1$ ?
Is the QCD evolution approximated correctly ?

Measurements on the genesis process of hadronic structures
use quantum fluctuations of the photon:
since over $20$ years we know the momentum distributions of quarks 
resulting from the photon splitting into quark-antiquark pairs.
For the first time, the HERA and LEP 
experiments have measured the gluon 
distribution of newly built hadronic configurations, which is found
to be very similar to the gluon distribution measured in protons.
The open question to the photon data is: is hadronic structure at low $x$
universal, i.e., do the low-$x$ partons ``know'' about the partons
in the high-$x$ region ?

Measurements of the partonic structure of colour singlet exchange
at HERA and the TEVATRON \cite{tevatron} for the first time show that 
such objects dominantly consist of gluons.
Will these measurements serve as a reference process for
a gluon driven regime and offer new insight into QCD dynamics ?

\vspace*{0.2cm}
Major contributions of lepton-hadron scattering in the past $10$ years
deepen our understanding of hadronic structures.
Burning open questions ensure that this field of research
will remain very active also in the coming decade.

\section*{Acknowledgments}
I wish to thank very much the Liverpool team for a wonderful conference !
For careful reading and comments to the manuscript I am grateful
to E. Elsen and B. Foster.
I wish to thank Th.~M\"uller and the IEKP group of the University
Karlsruhe for their hospitality, and the Deutsche Forschungsgemeinschaft 
for the Heisenberg Fellowship.

\end{document}